\documentclass[twocolumn,showpacs,preprintnumbers,amsmath,amssymb]{revtex4}
\usepackage{graphicx}
\usepackage{dcolumn}
\usepackage{bm}

\begin{document}

\title{ Lifshitz transitions in multi-band Hubbard models for topological superconductivity in complex quantum matter }
\author{Antonio Bianconi$^{1,2,3}$}

\affiliation{
$^1$RICMASS Rome International Center for Materials Science Superstripes, Via dei Sabelli 119A, 00185 Rome, Italy
\\
$^2$CNR-IC, Istituto di Cristallografia, Via Salaria Km 29.3, Monterotondo, Roma, I-00015, Italy
\\
$^3$National Research Nuclear University, MEPhI, Kashirskoe sh. 31, 115409 Moscow, Russia}

\date{31 Dec 2017, Journal of Superconductivity and Novel Magnetism DOI: 10.1007/s10948-017-4535-1}

\begin{abstract}

How the macroscopic quantum coherence can resist to the decoherence attacks 
of high temperature is a major challenge for the science of the 21st century. 
Superstripes 2017 conference held in Ischia on June 2017 has been focused on 
the new physics of high $T_c$  superconductors made of complex quantum matter.
Today the standard model of
high $T_c$ superconductivity which grabs the physics of complex quantum matter is
the multi-band Hubbard model where the dome of $T_c$ occurs 
by driving the chemical potential in the proximity of a topological Lifshitz transition.
The multi-gap superconductivity in the $T_c$ dome is driven by exchange interaction between
a first condensate in the BEC-BCS crossover which coexists with second BCS condensates.
The proximity to Lifshitz transitions in correlated electronic systems 
gives the ubiquitous arrested phase separation observed in all high temperature
superconductors. 
Non Euclidean filamentary hyperbolic geometry is needed for the space description of superstripes textures
produced by the coexistence of short range CDW puddles, 
hole poor SDW puddles and self organized dopants rich puddles. 
A road map to room temperature superconductors in particular organic compounds made of superlattices 
of quantum wires driven by Fano resonances with one of the condensates in the BEC-BCS crossover has been proposed. 
 
\end{abstract}


\maketitle
\section{Introduction.}

High temperature superconductivity has been found in a sequence of exotic complex materials: 
ceramics, inter-metallics, diborides, iron pnictides and chalcogenides, 
with the record for the highest critical temperature
$T_c$ in pressurized sulfur hydride near structural phase transitions.
These superconductors show a dome of high critical temperature
in a particular range of interstitials or defects concentration and in a particular range of pressure or misfit strain.
Each system shows a different complex landscape characterized by a different type
of multi-scale arrested phase separation with local-lattice-distortion, orbital, charge and spin modulations
forming textures of puddles of stripes from atomic-scale to nano-scale, mesoscale and micron-scale.
These exotic systems are clearly far away from a typical $conventional$ BCS superconductor 
made of homogeneous metal with a single large Fermi surface.

Experimental evidence for stripes due to anharmonic incommensurate lattice and orbital modulation has been 
found since 1990 by novel experimental methods using synchrotron radiation.
The early results have been reported and discussed 
at several international conferences focusing on lattice complexity
and phase separation \cite{1991-01a,1992-01a,1992-05x,1992-08x,1994-10a}.

The date of birth of the $stripes$ physics in high temperature superconductors can be fixed on Dec 7, 1993 
which is the priority date of the patent for material design
of heterostructures at atomic limit formed by superlattices 
of quantum stripes \cite{1993-12}. In this stripes scenario a Fermi liquid coexists with an incommensurate 1D 
charge density wave (CDW ) \cite{1994-07} forming a multigap superconductor near a Lifshitz transition where
the critical temperature amplification is driven by Fano resonances involving different condensates,
This stripes scenario for the high $T_c$ mechanism was presented at several international conferences in 1994-1996
\cite{ruani1995,tenyearsa}. 
The series of conferences on $stripes$ in high temperature superconductors started three years later on Dec 1996
 following the confirmation for the presence of stripes
by other standard experimental methods like neutron diffraction and 
NMR \cite{tenyears}, The first Stripes 1996 conference was 
followed by the second very large conference Stripes 1998 \cite{stripes98x} 
where the very simple Emery model of spin stripes 
with wave-vector $q_{sdw}$, locked with a charge stripes with wave-vector $q_{cdw}$= 2$q_{sdw}$ 
became very popular within the scientific community. 
The series of stripes conference have kept open the discussion on many different 
proposals for the stripes scenarios like 
the coexistence of short range charge stripes puddles unlocked from spin stripes puddles \cite{superstripes1999c}.
A major problem in the field was the diversity of the stripes scenarios in different families of hole doped cuprates 
which was solved in 2000 by the
disclosure of the key role of the lattice strain field with a critical strain value for the appearing of short range stripes ordering \cite{superstripes2000c}.
Moreover at the $stripes$ conference the term superstripes \cite{superstripes} has been coined to indicate 
the complex landscape generated by an arrested phase-separation near a critical strain point, 
forming a texture of  nano puddles of short range striped charge density wave order 
\cite{superstripes2000a,superstripes2000b,superstripes2000x,superstripes2001,superstripes2001a}.

 At that time, in the year 2000, the proposed heterogeneous landscape of superstripes was in contrast with the
 most popular paradigm  i.e., the single-band Hubbard model.
 Later many Scanning Tunneling Microscopy (STM) experiments have confirmed this scenario providing compelling 
 evidence for electronic nanoscale phase separation.
 Today the presence of nanoscale puddles of electronic pseudogap matter competing
 with superconducting condensate puddles with different symmetries \cite{erice2003} is well established.
In the new emerging paradigm high temperature superconductors 
 are described as particular cases of heterogeneous  complex quantum
 matter where particular forms of complexity are not detrimental (like normal disorder in the majority of disordered superconductors) but 
 they favor higher critical temperatures \cite{2005x}.

The name of the series of  $stripes$ conferences  changed its name into $superstripes$ conferences 
in 2008 driven by clear evidence 
for phase separation in iron based superconductors \cite{2009x,2009y}.
This decision marked the shift of the scientific interest toward 
quantum phenomena in complex matter \cite{super2011,super2012,super2013,super2014,super2015}.
and toward a new paradigm: the multigap superconductivity  near electronic 
topological Lifshitz transitions \cite{lif10} with pairing in one of the Fermi surface pockets is in the BEC-BCS crossover regime \cite{leggett} .

\section{From the  Woodstock of Physics to Superstripes 2017}

The international conference Superstripes 2017 has been held on June 4-10, 2017 at the Ischia island of the Neapolitan archipelago in Italy.
Scientists leaders in the field have been invited to discuss the latest advances in this field. 
Some discussions at Superstripes 2017 have given an answer to topics open since the 
APS March meeting held in New York, on 18 March 1987, 
the so called $Woodstock$ $of$ $Physics$, where
30 years ago Alex Muller presented the discovery
of high temperature superconductivity in ceramic La-Ba-Cu-O materials \cite{1}.
Paul C.W Chu reported superconductivity above liquid nitrogen temperature in $YBa_2Cu_3O_{6+y}$  ($Y123$) \cite{3}.
At Ischia 2017 conference Alex Muller, presented a review on the key role of complexity in cuprates recently  
reported in the book on high-$T_c$  copper oxide 
superconductors \cite{buss} presenting the scenario of high temperature superconductivity in complex materials.
Paul Chu discussed the role of interfaces in the enhancement of  $T_c$ above 77 K including the role of lattice architecture, 
internal strain and pressure. He pointed out the fact that a single-band Hubbard model 
is not sufficient for describing  high $T_c$,  why multi-band Hubbard models are needed \cite{chu}. 
Vladimir Kresin who proposed at the 1987 APS March 
meeting the theoretical scenario of strong  electron-phonon coupling in a complex lattice \cite{5} discussed in Ischia the  strong coupling 
limit in the pairing \cite{kres} involving high energy phonons
 in view of explaining 203 K superconductivity in pressurized $H_3S$ 
 which is today object of active research \cite{lifh3s-1,lifh3s-2,lifh3s-3,lifh3s-4}.

Band structure calculations of the parent compounds $La_2CuO_4$ and $YBa_2Cu_3O_6$ presented at the Woodstock of Physics 
predicted that the Fermi level is at half filling in a wide band due to 
the covalent bond between Cu(3d) [ml=2]  and O(2p$_{x,y}$) orbital in the metallic bcc $CuO_2$ layers.
Considering a perfect tetragonal bcc lattice
including only first-neighbor hopping t, the 2D Fermi surface at half filling is predicted
to have a square shape. The chemical potential is tuned
at the electronic topological Lifshitz transition
from the hole-like to the electron-like Fermi surface.
Here the system shows a peak in the Density of States (DOS) and strong electron-phonon scattering 
at the nesting wave-vector $2k_F$ connecting opposite sides of the square Fermi surface.
Therefore  high temperature superconductivity  was expected according with BCS theory
because of strong electron-phonon coupling and high density of states (DOS) at the van Hove singularity. 

On the contrary experiments have shown the failure of these predictions since
the parent compounds are antiferromagnetic insulators with an energy gap of about 2 eV. 
The insulating phase was explained by two different schools. 
The first school assumed the opening of 
a Peierls gap over the $full$ Fermi surface with wave-vector $2k_F$ 
associated with the formation of a 2D Peierls charge density wave.
The 2D Peierls CDW can be described as the ordering of polarons in the real space.
The CDW competes with 
the ordering of polaron pairs (bipolarons) in the k-space 
forming a superconducting phase in the strong coupling limit where below $T_c$ a Bose Einstein 
Condensation (BEC) occurs \cite{1981-02}.
The second  school proposed that the insulating phase was due to
the opening of a Hubbard gap $U_{dd}$ 
between [$3d^9$,$3d^9$]  and [$3d^8$,$3d^{10}$] configurations 
with the formation of a Mott insulator while the metallic superconducting phase 
was assigned to itinerant electronic singlets in a single band Hubbard model 
which condense like the preformed electron pairs in metal-ammonia solutions
 \cite{ogg} or like the resonating valence bond proposed by Pauling 
\cite{pauling} for transition metals.

\section{From single-band to multi-band Hubbard model and Lifshitz transitions}

It was well known that in the frame of a single-band Hubbard model the Mott insulator occurs if $U_{dd}$ is 
larger than the conduction band-width W. At the 1987 APS March meeting 
it was assumed by the scientific community that chemical doping form itinerant $Cu^{3+}$ impurity states,
with $3d^8$ configuration, moving in a background made of $Cu^{2+}$ ions with the $3d^9$ configuration.

Three weeks later on Apr 8th 1987 in the Symposium on High $T_c$ Superconductivity at the 7th General Conference 
of the Condensed Matter Division of the European Physical Society held in Pisa, Italy,
 it was reported \cite{1987-04} that the doped holes in doped 
 metallic $Y123$ do not form the expected $Cu^{3+}$ with
 $3d^8$ configuration but they form the 
 unexpected $Cu^{2+}$ $O^{1-}$ states called $3d^9L$, 
where L indicates the  hole in the ligand oxygen 2p orbital.

This result was obtained by Cu $L_{2,3}$-edge  x-ray absorption near edge structure (XANES) spectroscopy of the
high $T_c$ superconductor Y123  measured using ACO storage ring in Orsay, France in March 1987 \cite{1987-04,1987-09a} and 
by the Cu $K$-edge  XANES measured at the Adone storage
 ring in Frascati, Italy \cite{1987-09c}.

These results have been possible thanks to the development of XANES spectroscopy as a probe of 
both of multiple scattering resonances or shape resonances \cite{1978,1979,1980,1986-03}
and of many body electronic 
configurations  in valence 
fluctuation materials, heavy fermions and charge transfer correlated transition 
metal oxides like NiO, $CeO_2$ and $PrO_2$ \cite{1982-09,1985-02,1986-02,1987-01,1987-02,zaanen}.
                                                   
The  Cu $L_{2,3}$ X-ray photoelectron spectroscopy (XPS) spectra of Y123 were measured to get $U_{dd}$ in the Italian ENEA
Laboratory  \cite{1987-09d} and the results were found to be in agreement with the XPS experiments made by Fujimori in Tokyo \cite{1987-09e} 
at the same time.

These works have been confirmed in different cuprate families like in $La_{1.85}Sr_{0.15}CuO_4$ (La124)\cite{1988-02}.
The results have been presented at the four major international conferences in 1987:
1) the Pisa EPS April meeting \cite{1987-04}; 2) at the  Special Adriatico
Research Conference on High Temperature Superconductors held on 6-8 Jul 1987 in Trieste, Italy \cite{1987-09b}; 
3) at  the tenth Taniguchi international symposium held on October 19-23, 1987
in Kashikojima, Japan \cite{1987-09f};
and 4) at the 14th International Conference on x-ray 
and inner-shell processes, held in Paris on September 14-18, 1987  \cite{1987-10}.
The relevance of these results was recognized on 8 Dec 1987 at the Nobel prize ceremony
where Alex Muller \cite{2} reported that :
 
 [early photoelectron core-level spectra (XPS and UPS) by Fujimori 
  et al.\cite{1987-09e} and Bianconi et al.\cite{1987-09d}  in $La_{1-x}Sr_xCu_2O_{4-y}$ and
  $YBa_2Cu_3O_7$ did not reveal a final state owing
   to a $Cu^{3+}$ $3d^8$ state]
   
 In fact the  XPS experiments \cite{1987-09c,1987-09d} have clearly shown
that the parent compounds are Mott insulators with the Coulomb repulsion 
between two holes in the Cu 3d orbitals $U{dd}$= 6 eV  \cite{1987-09c,1987-09d}, larger than 
the conduction band width which was calculated by band structure calculations for non interacting fermions.
The Cu $L_3$ XANES experiments have shown that the carriers, created by doping,
are states with $3d^9 L$ many body configuration in the correlation gap.

The correlation gap in both La124 and Y123 compounds is not $U_{dd}$ as expected 
 for a single-band Hubbard model but the charge transfer gap for the excitation
 from the [$Cu(3d^9)$,$O(2p^6)$] to the  [$Cu(3d^{10})$,$O(2p^5)$] many 
 body configuration, called also the $3d^9$ to $3d^{10}L$ gap, which was predicted for the correlated charge transfer transition metal oxides
 \cite{zaanen}. 
 
A month after the EPS 1987 Pisa conference, Emery understood that the results of the 
Cu $L_3$-edge experiment  \cite{1987-04}
had falsified the single-band Hubbard model and proposed 
 in June 1987 the three-band Hubband model \cite{emery} involving p and d orbitals
for hole doped cuprates.

The experimental evidence of the  $3d^9L$ by A. Bianconi \cite{1987-09b} and  the V. Emery theory paradigm 
of the p-d three-band Hubbard model were presented together at the Adriatico Trieste meeting in July 1987.
 After the Adriatico 1987 meeting the fact that doped holes are in the oxygen orbital in cuprates was 
widely accepted by the community as recognized by Alex Muller in the opening talk at the 3rd international conference on 
Materials and Mechanisms of Superconductivity in 1991 at Kanazawa Japan \cite{2a} :

[The carriers in these type of conducting cuprates are holes on the oxygens. 
To me the first indication came from X-ray absorption spectroscopy by Prof. Bianconi at University of Rome] 

 Evidence for the $3d^9L$ states induced by doping in the correlation gap was 
presented in 1988 at the  first  $M^2S$ conference in Interlaken \cite{1988-06a,1988-06c,1988-06f}.
and at the international Symposium on the Electronic Structure of High $T_c$  Superconductors, 
Rome, 5-7 October 1988 \cite{1988-10bm}, 
where it was confirmed by many authors \cite{1988-01x,1988-04,1988-06d,1988-10a,1988-10b}. 
These results have been conformed by experiments and theories published
in 1989 \cite{1989-01,1989-09}; 
in 1990 \cite{1990-01,1990-07}; in 1991 \cite{1991-03,1991-11,1991-12}; 
in 1992  \cite{1992-04,1992-08}; 
in 1993 \cite{1993-05,1993-10}; 
in 1995-1997 \cite{1995-04,1997-04} 
and in these last 10 years \cite{2007-08,2010-08,2017-07}.

It was established that the correct model for high temperature superconductors 
is the multi-band Hubbard  model, where the domes of high $T_c$ 
occur in the proximity of electronic topological transitions called Lifshitz transitions.

Boguliobov equations for a multigap superconductor near a band edge have been solved by numerical calculations 
for the cases of sulfur hydrides, cuprates, diborides, and iron based superconductors \cite{lifh3s-1,lif1,lif2,lif3,lif4,lif5,lif6,lif8,lif9,lif11}. 
Recently it was shown that this mechanism for high $T_c$ can be in action in particular  filamentary organic  
compounds \cite{lifpter} where the condensate in the new appearing Fermi surface is in the BEC-BCS crossover regime.

Also the ubiquitous nanoscale phase separation  occurring 
in high temperature superconductors has been shown to be driven
 by the proximity to a Lifshitz transition in multi-band Hubbard models \cite{kus,kug1,kug2,kug3}.

\section{Recent advances}

At superstripes 2017 conference it was proposed that the origin of 
the multiple Fermi surfaces could be determined by oxygen interstitials self-organization 
as it has been shown  in the case of HgBa$_2$CuO$_{4+y}$ (Hg1201)  where the oxygen interstitials (O-i)  
are not homogeneously distributed but form one-dimensional atomic wires which can dramatically enhance the critical temperature. \cite{jar}.
The complex interplay of charge, spin \cite{tall,palu,cha,yana} and orbital degrees
 of freedom including spin-orbit interaction in topological matter \cite{ole,glo} 
 are now object of investigation as key ingredients. 
 The rich physics of unconventional superconductivity in cuprates and
their asymmetry between electron doped and hole doped families continue to attract theory and experimental research
\cite{mar,tei,iva,mosk,kan,zhao}. Ivanov has shown the noncentrosymmetric structure of a Bi2212 crystal at optimum doping
 in a high magnetic field by using x-ray magnetic circular dichroism, XMCD.
This new result suggests a possible role of spin-orbit coupling. 
Investigation of  basic physics in low dimensional quasi 1D 
superconductors \cite{pto,bar} and in the 2D electron gas  
\cite{ala,pud} have been object of high interest to unveil key features of dimemsionality in unconventional superconductors. 
Iron based superconductors continue to provide a clear 
case for the study of the interplay of nanoscale phase separation and multi-gap superconductivity \cite{duan,umm,shy}. 
New topics in physics and materials science are provided 
by systems out-of-equilibrium \cite{tod}, novel silver based materials \cite{groc} and granular materials \cite{mos}. 
Finally key advances in the quantum electronics using superconducting devices \cite{sem,gala} and on metal-to-insulator transition in $VO_2$ \cite{marcelli} have been reported at Superstripes 2017 

\section{Conclusions}

There is today after many years of discussion a growing agreement 
in the scientific community on some key common physical features of high temperature superconductors. 
All  high $T_c$ superconductors are highly inhomogeneous and different
multi-band Hubbard models are needed to describe unconventional high temperature superconductors.
Topology, Lifshitz transitions, and spin-orbit coupling are opening new fields of research in condensed matter
 of strongly correlated materials. Further works will be addressed to exchange interaction giving Fano
  resonances between condensates in the strong coupling regime in hot spots of the Fermi surface 
  coexisting with condensates in other portions of the Fermi surfaces in the weak coupling regime, which can be manipulated in 
  metalorganic materials to get room temperature superconductors \cite{lifpter}.

\end{document}